%% file: 1-main.tex
\documentclass[twocolumn,a4paper,10pt]{article}
\input{preamble.tex}
\title{Photodissociation of \TheComplex{}: A non-adiabatic dynamics investigation}
\author{Bartosz Ciborowski, Morgane Vacher\thanks{morgane.vacher@univ-nantes.fr} \\
        Nantes Universit\'{e}, CNRS, CEISAM UMR 6230, F-44000 Nantes, France}
\date{January 10, 2025}
\begin{document}
\twocolumn[
    \maketitle
    \begin{onecolabstract}
        \input{Abstract.tex}
        \vspace{2em}
    \end{onecolabstract}
]
\saythanks
\section{Introduction}
	\input{Introduction.tex}
\section{Methods}
    \input{Methods.tex}
\section{Results and Discussion}
    \input{Results.tex}
\section{Conclusions}
    \input{Conclusions.tex}
\section{Acknowledgements}
    \input{Acknowledgements.tex}
    \printbibliography
    \newpage
\end{document}


\maketitle
    \tableofcontents
    \listoffigures
    \listoftables
    \newpage
\section{Franck--Condon region Benchmark}
TD-DFT and CASSCF methods were benchmarked at B3LYP/6-31G* optimized geometry with a $C_{2v}$ symmetry constraint (coordinated are present in Table \ref{Tab:B3LYP-geom}). Results of the benchmarked along with reference data\autocite{Guillaumont1997,Guillaumont2001} are shown in Table \ref{Tab:FC-Benchmark-SI}. The functionals used are B3LYP, PBE0, BHHLYP, CAM-B3LYP. Dependence of TD-DFT excitation energies on chosen basis set is shown in Table \ref{Tab:FC-Benchmark-2-SI}. These tables are not exhaustive, as the character of high-energy TD-DFT states could not be unambiguously assigned due to their composition of many low-contribution orbital transitions, and a complete state characterisation was not the objective of this study. B3LYP and PBE0 3$d$$\rightarrow$$\pi^*_{\text{bpy}}$ excitations show charge-transfer failure,\autocite{Peach2008} manifested through underestimation of energy when compared to reference data and functionals with more, or coulomb-attenuated, exact exchange. In a reverse fashion, BHHLYP consistently overestimates 3$d$$\rightarrow$$\pi^*_{\text{bpy}}$ energies by 0.2 -- 0.3 eV when compared to CAM-B3LYP, while giving very similar 3$d$$\rightarrow$$\pi^*_{\text{CO}}$ energies.

\begin{figure}[h]
    \centering
    \includegraphics{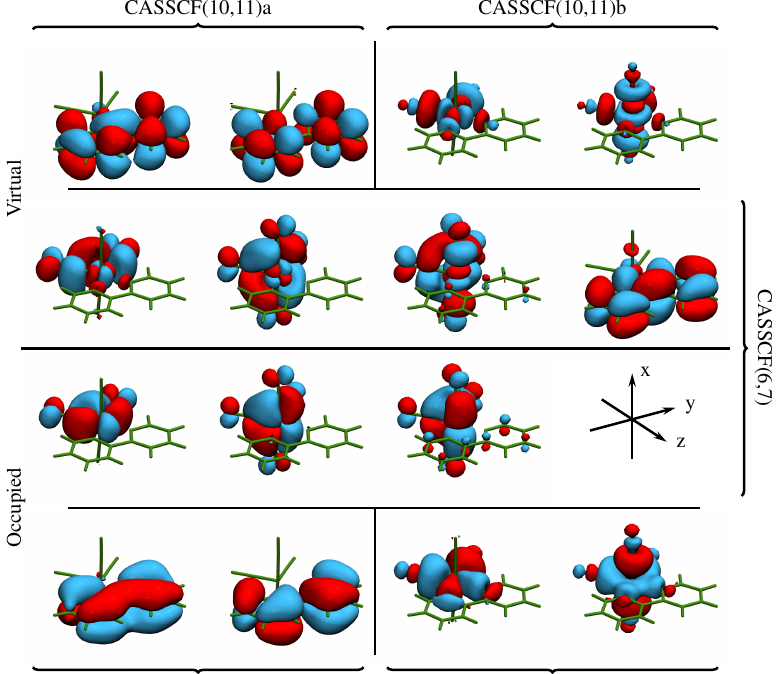}
    \caption[CASSCF Orbitals]{Orbitals of active spaces using during the study. CASSCF(6,7) is made up of the middle row of orbitals. CASSCF(10,11) (a) and (b) extend on top of CASSCF(6,7) orbitals with 2 pairs of bonding/antibonding orbitals. CASSCF(10,11)a adds orbitals on the left; CASSCF(10,11)b adds orbitals on the right.}
    \label{Fig:CAS-Orbs-SI}
\end{figure}

Though 3$d$$\rightarrow$$\pi^*_{\text{CO}}$ transitions are formally of MLCT character, they do not show charge transfer failure like 3$d$$\rightarrow$$\pi^*_{\text{bpy}}$ transitions do. 3$d$ orbitals involved in these excitations are more precisely pairs of bonding/antibonding orbitals between chromium and neighbouring carbonyls. This can be seen in Figure \ref{Fig:CAS-Orbs-SI}, which showcases orbitals used in CASSCF methods, which are analogous to TD-DFT orbitals. As such, 3$d$$\rightarrow$$\pi^*_{\text{CO}}$ transitions have large orbital overlap, and therefore are local in character and do not suffer from charge-transfer failure. As a consequence, CAM-B3LYP is in closest agreement with MRCI/CASSCF values, with BHHLYP being second (Mean Average Error: 0.175 and 0.337 eV).
\input{Tab-FC-benchmark.tex}
\input{Tab-FC-B3LYP-xyz.tex}
\FloatBarrier
\newpage
\section{Active spaces of CASSCF calculations}
\FloatBarrier
\newpage
\section{Further benchmark on Cr-CO$_{\text{ax}}$ PES}
\begin{figure}[h]
    \centering
    \includegraphics{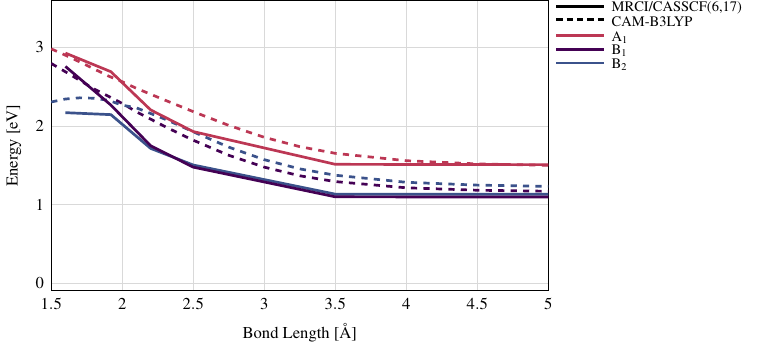}
    \caption[PES of CAM-B3LYP Excitation Energies]{Excitation energies of CAM-B3LYP and MRCI/CASSCF(6,17)\autocite{Guillaumont2001} as a function of Cr--CO\textsubscript{ax} bond length.}
    \label{Fig:RelativePES-SI}
\end{figure}
\begin{figure}[h]
    \centering
    \includegraphics{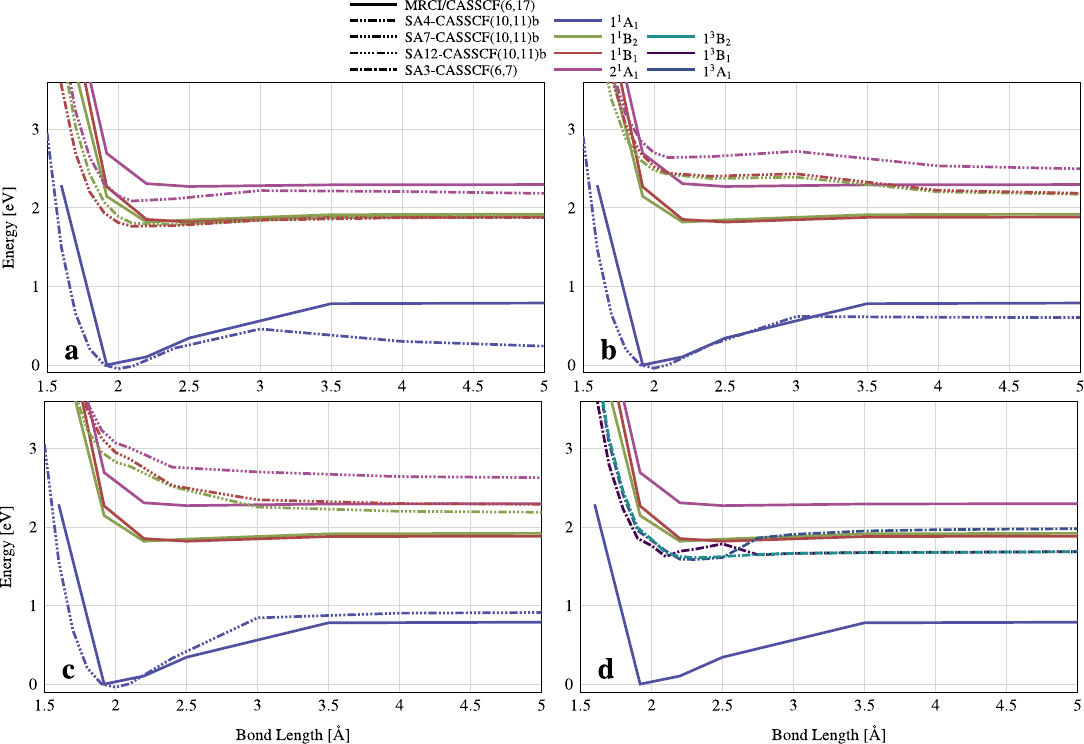}
    \caption[PES of CASSCF(10,11)b and triplet CASSCF(6,7) methods]{Potential energy surfaces of \TheComplex{} as a function of Cr--CO$_{\text{ax}}$ bond length coordinate; (\textbf{a}) SA4-CASSCF(10,11)b; (\textbf{b}) SA7-CASSCF(10,11)b; (\textbf{c}) SA12-CASSCF(10,11)b; (\textbf{d}) Triplet SA3-CASSCF(6,7); Reference MRCI data reproduced from Guillaumont \emph{et al}.\autocite{Guillaumont2001}; PES scans were initiated from a B3LYP/6-31G* minimum geometry; State labels correspond to the symmetry at $C_{2v}$ FC geometry.}
    \label{Fig:CAS-PES-SI}
\end{figure}
\begin{figure}[h]
    \centering
    \includegraphics{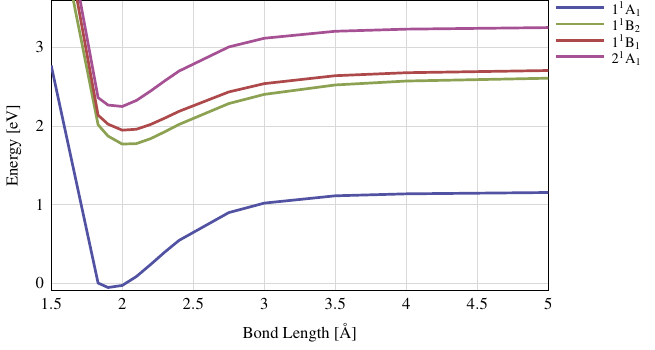}
    \caption[CASSCF(6,7) PES of an equatorial carbonyl]{Potential energy surface of \TheComplex{} as a function of Cr--CO$_{\text{eq}}$ bond length coordinate using SA4-CASSCF(6,7); PES scan was initiated from a B3LYP/6-31G* minimum geometry; State labels correspond to the symmetry at $C_{2v}$ FC geometry.}
    \label{Fig:CAS-PES-COeq-SI}
\end{figure}
\begin{table}[h]
    \centering
    \caption[Contribution of 3$d_{x^2}$ basis function on axial PES]{Contribution of the 3$d_{x^2}$ basis function to the $\pi^*_{\text{bpy}}$ orbital solution as a function of Cr--CO bond length.}
    \label{Tab:dx2-SI}
    \input{Tab-dx2-contribution-SI.tex}
\end{table}
\FloatBarrier
\newpage
\section{Dynamic calculations}
\input{Tab-FC-CASSCF67-xyz.tex}
\begin{figure}[h]
    \centering
    \includegraphics[width=\linewidth]{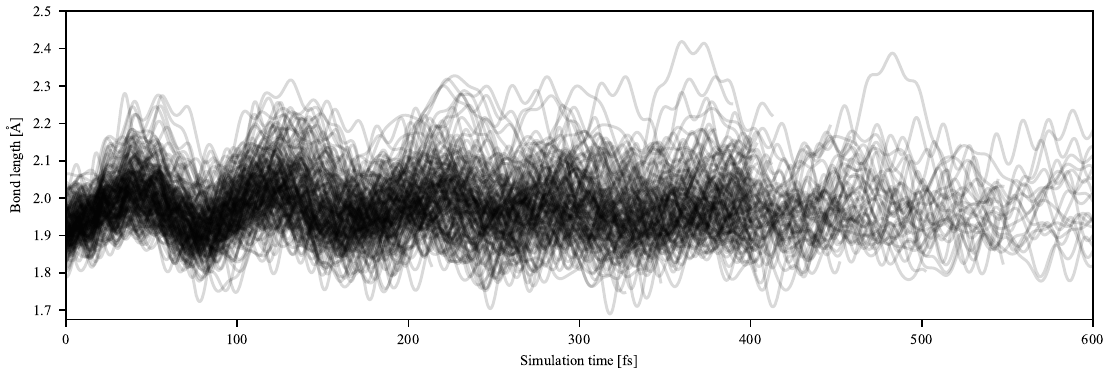}
    \caption[Bond length of all Cr--CO$_{\text{eq}}$]{Evolution of the Cr--CO$_{\text{eq}}$ bond length of all trajectories.}
    \label{Fig:BondLengthsEq}
\end{figure}
\begin{sidewaysfigure}
    \centering
    \includegraphics[width=.8\linewidth]{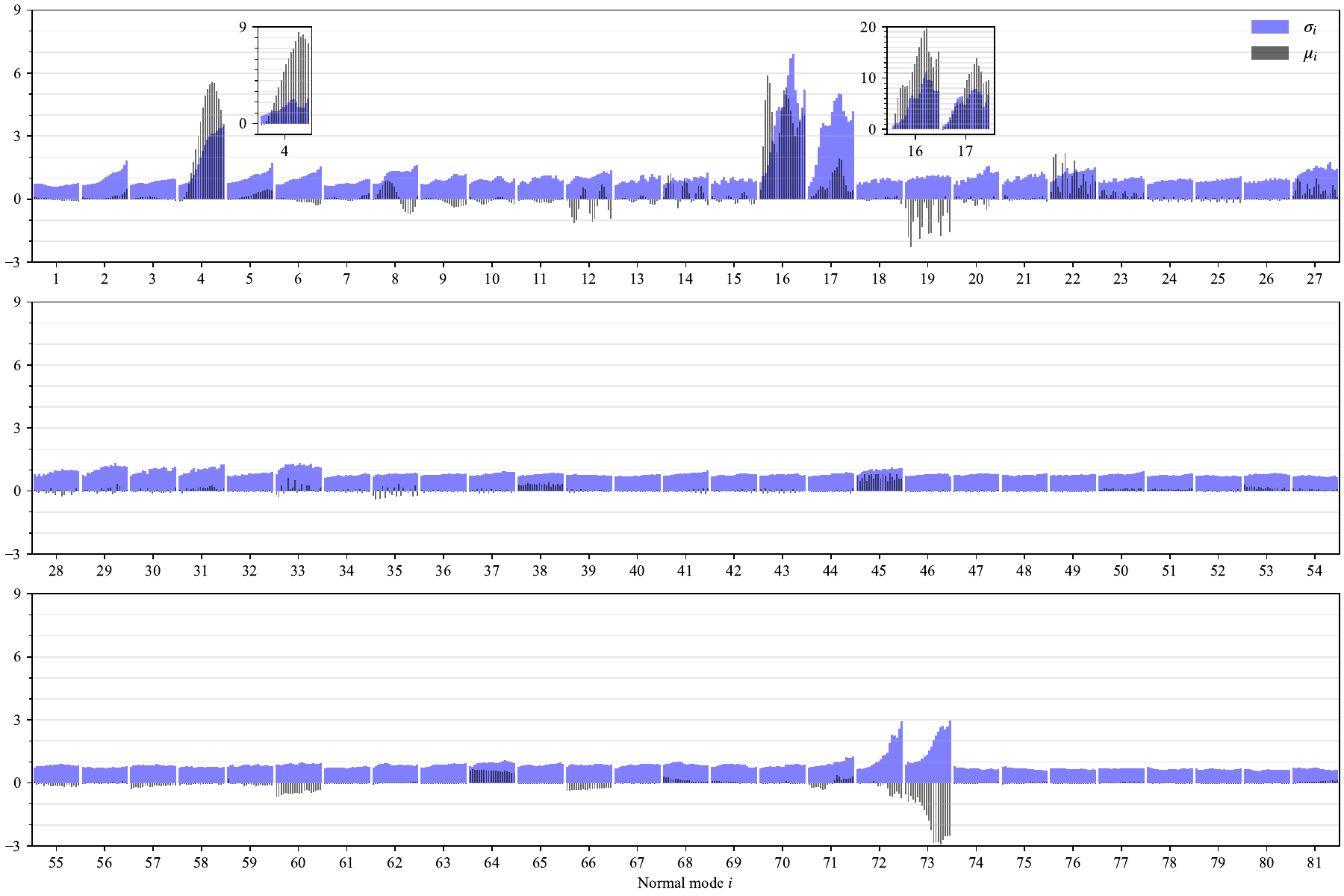}
    \caption[Time evolution of normal modes]{Mean displacement $\mu_i$ and standard deviation $\sigma_i$ for each mass-frequency scaled normal mode $i$ of the trajectory ensemble, taking the ground state minimum geometry as reference. Average values were taken over each 20 fs interval, up to the final value of 400 fs. Insets present the same analysis for a subset of dissociating trajectories, for modes 4, 16 and 17.}
\end{sidewaysfigure}
\begin{sidewaysfigure}
    \centering
    \includegraphics[width=.8\linewidth]{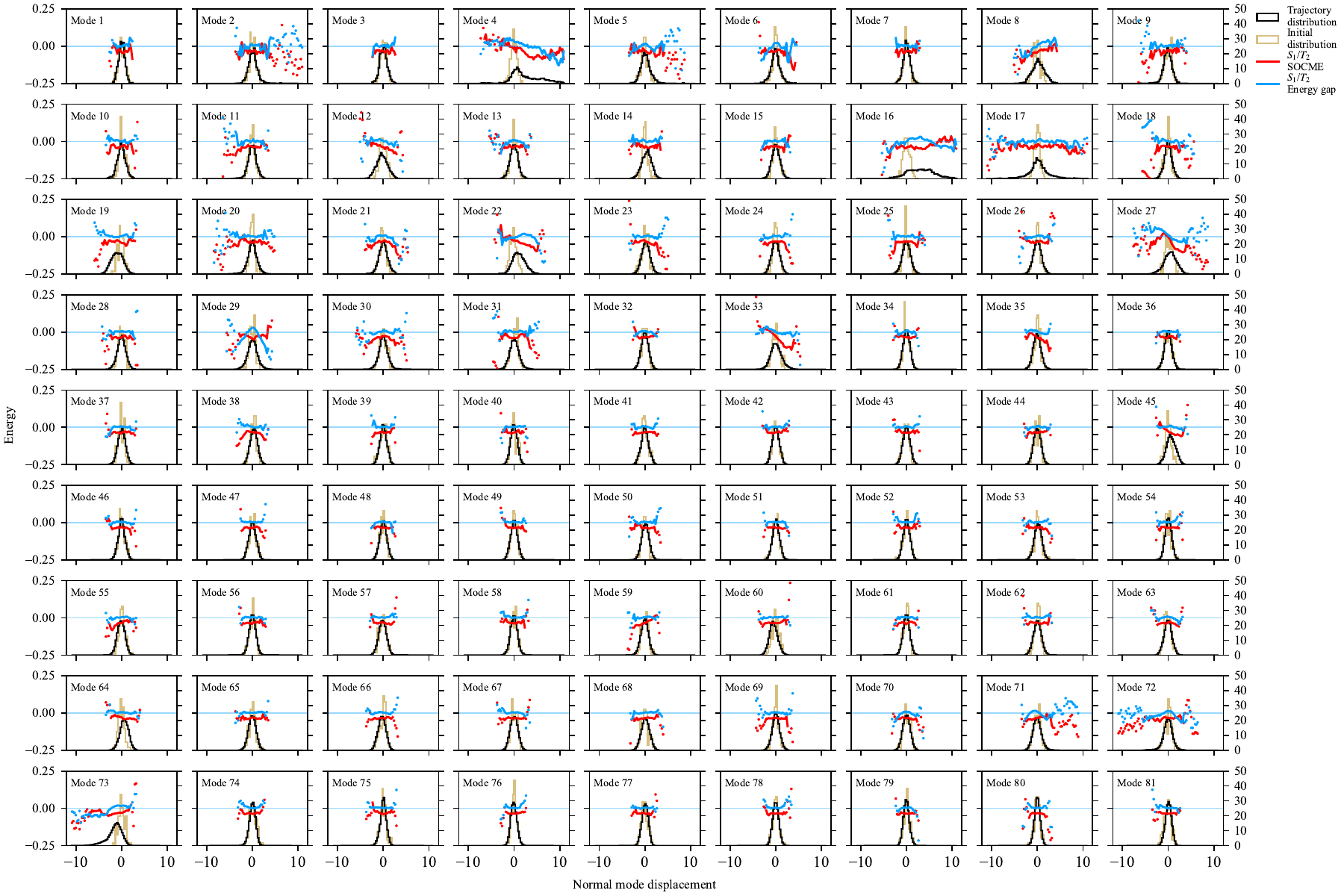}
    \caption[Dependence of $S_1$/$T_2$ energy gap and SOCME on normal mode displacement]{Energy gap and spin-orbit coupling matrix elements of $S_1$/$T_2$ crossing, averaged over the nuclear displacement for all normal modes. Distribution of all geometries visited during dynamics, and the initial Wigner distribution are given for reference. Energy gap values are given in eV (left axis), SOCME are given in cm$^{-1}$ (right axis). Areas with under 100 samples are indicated with dots.}
\end{sidewaysfigure}
\FloatBarrier
\printbibliography

%% file: preamble.tex
\usepackage[margin=2cm]{geometry}
\usepackage[T1]{fontenc}
\usepackage[backend=biber,style=chem-rsc]{biblatex}
\usepackage{graphicx}
\usepackage[table]{xcolor}
\usepackage{amsmath}
\usepackage{amssymb}
\usepackage{mathtools}
\usepackage{mathptmx}
\usepackage{array}
\usepackage{textgreek}
\usepackage{indentfirst}
\usepackage{caption}
\usepackage{multicol}
\usepackage{booktabs}
\usepackage{abstract}

\usepackage{hyperref}
\hypersetup{
	hidelinks=true,
	colorlinks=true,
	linkcolor=blue,      
	citecolor=blue
}
\newcommand{\TheComplex}{Cr(CO)$_4$bpy}

%% file: Abstract.tex
Carbonyl complexes of $d^6$ metals with an \textalpha-diimine ligand exhibit both emission and ligand-selective photodissociation from MLCT states. Studying this photodissociative mechanism is challenging for experimental approaches due to an ultrafast femtosecond timescale and spectral overlap of multiple photoproducts.
The photochemistry of a prototypical system \TheComplex{} is investigated with non-adiabatic dynamic simulations. Obtained 86 fs lifetime of the bright $S_3$ state and 13\% quantum yield are in good agreement with experimental data. The present simulations suggest a ballistic mechanism of photodissociation, which is irrespective of the occupied electronic state. This is in contrast to the previously established mechanism of competitive intersystem crossing and dissociation. Selectivity of axial photodissociation is shown to be caused by the absence of an avoided crossing in the equatorial direction.

%% file: Introduction.tex
Metal complexes with a $d^6$ electron configuration exhibit rich excited state behaviour through a combination of an electron-rich centre, closely spaced d-orbitals and a shell of stabilizing and electron-accepting ligands.
One of their most intriguing properties is the ligand-selective photodissociation under UV light, for instance of a carbonyl ligand~\cite{Trushin1998}. Following this process, a vacant site is formed on the metal's coordination shell, allowing for the association of a substrate. This property has been used in photocatalysis for various C--C bond forming and breaking catalytic cycles~\cite{Szymanska1997,Szymanska2006}.

When combined with electron-accepting \textalpha-diimine ligands, such as bipyridine (bpy) or phenantroline (phen), these complexes exhibit strong luminescence from metal-to-ligand charge transfer (MLCT) states. This property has been studied and used extensively in Re(I)~\cite{Striplin2001,Lo2010}, Ru(II)~\cite{Juris1988,McFarland2005,Cannizzo2006,Chou2006}, Os(II)~\cite{Chou2006}, Ir(III)~\cite{Nazeeruddin2003,Ho2020} based complexes~\cite{Evans2006,Moreira2010}. Ru(bpy)$_3^{2+}$ in particular is one of the most studied lumophores, but despite this, its electronic structure of the excited state remains a point of contention~\cite{Dongare2017}, highlighting the complexity of transition metal systems.

A metal core with strong spin-orbit coupling with a flexible ligand shell allows for design of versatile and tunable systems. However, rates and quantum yield of intersystem crossing are often insufficiently described by the central heavy-atom spin-orbit coupling alone~\cite{Chergui2012,Penfold2018}. The interaction between nuclear, electronic and spin degrees of freedom can lead to an intersystem crossing markedly faster~\cite{Bram2012} or slower~\cite{vanderVeen2011} than the atomic spin-orbit coupling would predict. Ultrafast processes can also cascade to give rise to such properties as light-induced spin-crossover~\cite{Decurtins1984,Cannizzo2010,Papai2016,Papai2022}. The electronic excitation may also be localized on one of the aromatic ligands and mediated by interligand charge transfer processes~\cite{Yeh2000,Rimgard2018,Perrella2023}, which are of major importance when the complex is bound to a superstructure, such as in light-harvesting devices~\cite{Mayo2006,Bomben2009}. Theoretical studies of these systems are hence necessary, but challenging.

Group 6 complexes of the structure M(CO)$_4$(\textalpha-diimine) [M = Cr, Mo, W], exhibit both luminescence and photodissociation pathways from MLCT states, and have been studied less so for applications, but rather as model systems for spectroscopic and theoretical approaches~\cite{Wrighton1975,Balk1980,Manuta1986,Kalyanasundaram1988,Wieland1990,Wen-Fu1996,Wen-Fu1997,Wen-Fu1998,Farrell2001,Vlcek2002,Rohrs2019}. The simplest of these, \TheComplex{} (shown on Figure \ref{Fig:PES}b), has received most attention and is the focus of this study as well~\cite{Vichova1992,Virrels1996,Guillaumont1997,Farrell1999,Guillaumont2001,Farrell2002,Ma2016}. While photodissociation is possible from complexes of above-mentioned metal cations, the combination of increased spin-orbit coupling for heavier metals~\cite{Vlcek2002} and choice of ligand can forbid or change the nature of the photodissociation pathway. In general, it occurs most prominently in complexes with both an \textalpha-diimine ligand and a carbonyl in an axial position to the \textalpha-diimine ligand. Through molecular design, dissociation of alkyl ligands~\cite{Gabrielsson2006} and from equatorial positions~\cite{Gabrielsson2004,Gabrielsson2008} have also been achieved.

M(CO)\textsubscript{4}(\textalpha-diimine) systems exhibit dual emission~\cite{Servaas1985,Stufkens1990} and wavelength-dependent quantum yield of photodissociation of an axial carbonyl~\cite{Wrighton1975,Vichova1992,Balk1980,Manuta1986} in two energy regimes: \emph{low} Cr$\rightarrow$\textalpha-diimine, and \emph{high} Cr$\rightarrow$CO MLCT states.
The \emph{high}-energy regime is present in M(CO)$_6$ complexes, and its photodissociation is the process extensively used in photocatalysis, while the \emph{low}-energy regime is introduced by the \textalpha-diimine ligand, and as such, its properties are sensitive to the choice of ligand and bonding to the metal. Photodissociation reactivity in the \emph{low}-energy regime is attributed to an avoided crossing of the low 3$d$$\rightarrow$$\pi^*_{\text{bpy}}$ $^1$MLCT state with a higher lying 3$d$$\rightarrow$3$d_{x^2}$ state, shaping the lower state to a quasi-bound or dissociative shape~\cite{Guillaumont2001,Vlcek2002,Rosa1996,Finger1995-2}. The strong wavelength dependence of dissociation quantum yield is understood to come from the transition to higher vibrational states, which facilitate passing over the energy barrier formed by the avoided crossing~\cite{Vlcek2002}.

Two trapping triplet states with the lifetimes of 8 and 87 ps were identified for \TheComplex{} through time-resolved experiments, to be the main photoproduct, when excited in the \emph{low}-energy regime~\cite{Virrels1996,Farrell1999,Farrell2002}. They are understood to be the source of emission and in direct competition with the photodissociation pathway. With picosecond triplet lifetimes, their depopulation is the limiting step for the femtosecond dissociation process. Conversely, if luminescence is of interest, then photodissociation is an undesirable substrate-depleting process.

Previous experimental works by Farrell \emph{et al.}~\cite{Farrell1999,Farrell2002} have reported that photodissociation completes within 400 fs. Later work by Ma \emph{et al.}~\cite{Ma2016} has shown, through better temporal resolution, that this process finishes within 100 fs. Exact photodissociation rate could not be obtained due to spectral overlap of absorption with triplet states and a low quantum yield, so an estimate of ($\sim$2.3 ps)$^{-1}$ was given, based on an assumption of competitive branching between photodissociation and intersystem crossing to $^3$MLCT states. The quantum yield of photodissociation for the \emph{low}-energy 3$d$$\rightarrow$$\pi^*_{\text{bpy}}$ states ranges from 1\% to 10\%, depending on experimental conditions~\cite{Balk1980,Vichova1992}.

To the best of our knowledge, the only previous theoretical work to use dynamic methods on \TheComplex{} is that of Guillaumont \emph{et al.}~\cite{Guillaumont1997,Guillaumont2001}.
They calculated MRCI/CASSCF PES, which we use as reference data, and propagated a nuclear wavepacket in one dimension.
We aim to expand on insights of that study by using non-adiabatic simulations in full dimensionality, using the surface hopping method~\cite{Tully1971,Tully1990}.

Our objectives are to obtain nuclear dynamic information and use it to refine the mechanism of this ultrafast photodissociation process by determining the origins of ligand selectivity and establishing the extent of competition between photodissociation and intersystem crossing.

In section 2, we present the theoretical methods used in the present study. In section 3, we begin by reporting the results of benchmarking electronic structure methods along the Cr--CO$_{\text{ax}}$ stretching coordinate, followed by the results of surface hopping dynamic simulations. We then discuss the obtained results in context of previous literature.

%% file: Methods.tex
All calculations in this work were performed in gas phase.
The equilibrium geometry of the electronic ground state was optimised at B3LYP/6-31G$^*$ level of theory (structure shown in Table S3) in OpenMolcas~\cite{MOLCAS}, under $C_{2v}$ symmetry constraint. It has been shown previously to give good Cr--C and C--O bond lengths in Cr(CO)$_6$~\cite{Villaume2007}. TD-DFT excitation energies with B3LYP, PBE0, BHHLYP, CAM-B3LYP functionals and 6-31G$^*$, 6-311G$^*$, 6-31+G$^*$, 6-311+G$^*$ basis sets were calculated at the B3LYP/6-31G* minimum geometry in ORCA 5.0.3~\cite{ORCA5} (Tables S1 \& S2) to evaluate the accuracy of TD-DFT and its sensitivity to chosen functional and basis. 

Next, rigid potential energy scans along the Cr--CO$_{\text{ax}}$ bond stretching coordinate were calculated using HF/CIS, MP2 and TD-DFT, compared against MRCI/CASSCF reference data~\cite{Guillaumont2001} (Figure \ref{Fig:PES}a). The reference data features a CASSCF(6,17) ground state and MRCI/CASSCF(6,17) excited states: a CASSCF(6,17) solution with a CISD calculation on each reference with a contribution of >0.08 to the CASSCF state. Due to poor performance of single-reference methods, calculations were performed with State Averaged CASSCF in OpenMolcas (Figures \ref{Fig:PES}b and S3). Several active spaces were attempted, but due to active space- and symmetry-breaking issues, only five CASSCF scans were completed successfully. The active spaces were composed of: (1) AS(6,7) - three pairs of metal bonding/antibonding orbitals (3$d_{xz}$, 3$d_{z^2-y^2}$, 3$d_{xy}$), plus first $\pi^*$ orbital of bipyridine; (2) AS(10,11)a - additional two pairs of $\pi$/$\pi^*$ bipyridine orbitals; (3) AS(10,11)b - adding remaining metal orbital pairs to AS(6,7) (3$d_{x^2}$, 3$d_{yz}$) (see Figure S1). CASSCF solutions were calculated using 4 states for AS(1) and 12 states for AS(2); AS(3) was calculated with 4, 7 and 12 states, giving five methods SA4-CASSCF(6,7), SA12-CASSCF(10,11)a, SA4-CASSCF(10,11)b, SA7-CASSCF(10,11)b and SA12-CASSCF(10,11)b. All CASSCF calculations were performed with the ANO-RCC-VDZ basis set.

SA4-CASSCF(6,7) was selected for use in surface hopping~\cite{Tully1971,Tully1990} calculations. Four singlet and three triplet states were used as active states, and so triplet SA3-CASSCF(6,7) PES were additionally calculated, along with SA4-CASSCF(6,7) CO$_{\text{eq}}$ PES (shown in Figures S3d and S4, respectively). Dynamics calculations were performed using SHARC 3.0 software,~\cite{Mai2018,SHARC3.0} interfaced with OpenMolcas. A 0K Wigner distribution of 100 geometries and velocities was randomly generated at the SA4-CASSCF(6,7) minimum geometry (structure shown in Table S5) using SHARC. All sampled geometries were initiated on the $S_3$ state, which is the bright 3$d_{xz}$$\rightarrow$$\pi^*_{\text{bpy}}$ state. Trajectories were propagated with local diabatization~\cite{Granucci2001}, energy-based decoherence~\cite{Granucci2007,Granucci2010} (empirical parameter of 0.1 au), in the diagonal representation, using a timestep of 0.5 fs. Two electronic structure calculations were performed at each timestep, SA4-CASSCF(6,7) for singlet states and SA3-CASSCF(6,7) for triplets states, giving state energies and gradients for all 7 considered states, both using the ANO-RCC-VDZ basis set.
Spin-orbit coupling effects were included through the Douglas-Kroll Hamiltonian to second order.

Two criteria of energy conservation were used to stop trajectories: (1) a deviation by 0.3 eV of total energy from the beginning of trajectory; (2) a step of 0.25 eV total energy during a single timestep. Additionally, individual state energies of all trajectories were monitored for breaking down of the active space. Trajectories for which the active space has broken were stopped, regardless of their total energy. These trajectories contribute to all data analyses up until the failure point, after which the data is renormalized, if required. Majority of stopped trajectories were caused by the breaking of the active space; in total 20 out of 100 at 300 fs and 37 out of 100 at 400 fs. All remaining trajectories were calculated up to 400 fs, which is taken as the simulation end point for the ensemble, but several trajectories were continued beyond to 600 fs in order to investigate nuclear evolution on longer timescales.

%% file: Results.tex
First presented are PES along the Cr--CO$_{\text{ax}}$ bond stretching coordinate for studied methods (initial benchmark of excited state methods at the Franck-Condon (FC) region is presented in SI). Then, results of surface hopping dynamic simulations are shown and discussed, split into nuclear and electronic dynamics. Finally, insights from combined electronic and nuclear dynamics are presented. 

\begin{figure*}[h]
    \centering
    \includegraphics[width=.8\linewidth]{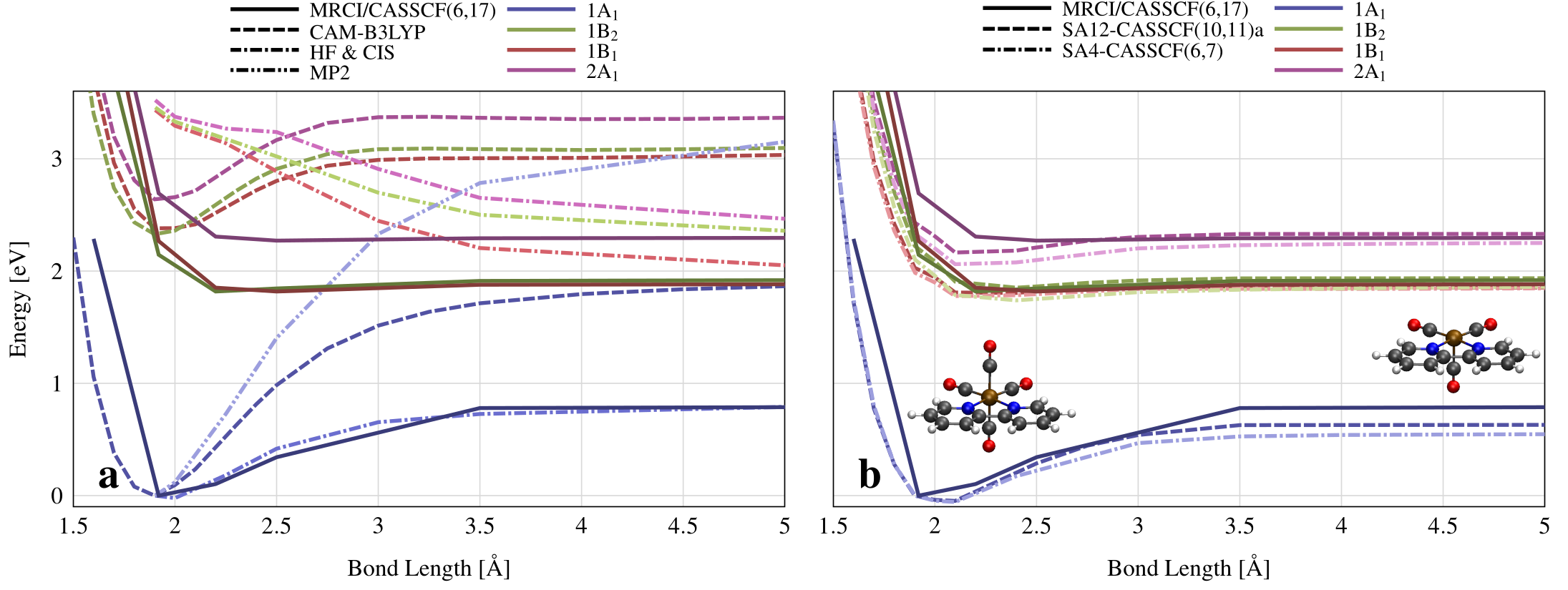}
    \caption{Singlet potential energy curves as a function of Cr--CO$_{\text{ax}}$ bond length; (\textbf{a}) Single reference methods; (\textbf{b}) SA-CASSCF methods; Reference MRCI/CASSCF(6,17) data reproduced from Guillaumont \emph{et al}.~\cite{Guillaumont2001}; PES scans were initiated from a B3LYP/6-31G* minimum geometry; orbitals included in each active space are detailed in text. State labels correspond to the symmetry at $C_{2v}$ FC geometry.}
    \label{Fig:PES}
\end{figure*}

\paragraph*{Potential Energy Scans}
Potential energy curves of single-reference methods are compared against high-level MRCI/CASSCF(6,17) data in Figure \ref{Fig:PES}a.
All studied states are of 3$d$$\rightarrow$$\pi^*_{\text{bpy}}$ MLCT character. The highest singlet $S_3$, denoted 2A$_1$ as per its FC symmetry, is a 3$d_{xz}$$\rightarrow$$\pi^*_{\text{bpy}}$ state. The lower $S_2$ and $S_1$ states, denoted 1B$_1$ and 1B$_2$, are of 3$d_{z^2-x^2}$$\rightarrow$$\pi^*_{\text{bpy}}$ and 3$d_{xy}$$\rightarrow$$\pi^*_{\text{bpy}}$ characters, respectively. See Figure S1 in ESI for orbital graphical representations.
The high level MRCI/CASSCF(6,17)~\cite{Guillaumont2001} curves feature a ground state barrier of $\sim$0.8 eV and excited states show shallow minima. All single reference methods fail to qualitatively capture the ground state shape of PES. HF follows reference ground state very closely, not showing any failure. UHF and UKS calculations of several functionals all give the spin operator value $\left<S^2\right>$ of 0.0 along the entire reaction coordinate, showing that a ground state dissociation produces two singlet products, and is of singlet character throughout. The ground state of CAM-B3LYP has a significantly overestimated energy barrier of $\sim$1.9 eV. Its transition energies at FC and dissociation-limit geometries are in excellent agreement with MRCI/CASSCF(6,17), but intermediate geometries give overestimated energies of up to 0.5 eV (see Figure S2). Irrespectively, the erroneous shape of the ground state renders all excited states non-dissociative. The same PES calculations were repeated with all DFT functionals available in ORCA 5.0.3 (not shown), to investigate if this is a functional or method dependent problem. All have produced the same overestimated ground state energy barrier, showing that DFT is fundamentally unable to describe this bond-breaking process and that a different method needs to be used. MP2 shows typical failure of describing dissociation processes, based on which ADC(2) calculations were not attempted. These considerations have lead us to consider CASSCF as the method of choice.

Potential energy curves of SA4-CASSCF(6,7) and SA12-CASSCF(10,11)a are shown in Figure \ref{Fig:PES}b. All CASSCF variants qualitatively reproduce the shape of reference curves. SA4-CASSCF(6,7) and SA12-CASSCF(10,11)a show minimum ground state geometries bond lengths of 2.1 \AA{} (1.9 \AA{} for MRCI/CASSCF(6,17) and B3LYP) and excited state energy barriers: $\sim$0.1 eV for S$_1$ and S$_2$; $\sim$0.2 eV for S$_3$ (0.1 eV and 0.02 eV for MRCI/CASSCF(6,17)). Interestingly, neither method explicitly includes orbitals or states involved in avoided crossings between 3$d$$\rightarrow$3$d_{x^2}$ and lower excited states, which are well established in literature to be the cause of quasi-bound excited state shapes~\cite{Vlcek2002}. Reference MRCI/CASSCF(6,17) data does not include the relevant 3$d_{x^2}$ orbital in its active space either, but this interaction is expected to be incorporated through the post-CASSCF CISD calculation (although not for the ground state, which is just CASSCF(6,17)).
Analysis of the CASSCF solution shows that, as the bond stretches, the $\pi^*_{\text{bpy}}$ orbital gains a contribution from the 3$d_{x^2}$ basis function (see Table S4). CASSCF(10,11)b calculations (Figure S3a-c), which explicitly include the relevant $d_{x^2}$ orbital at FC geometry, show a much greater contribution of 3$d_{x^2}$ basis function as the bond stretches, clearly showing state mixing. Points at which contribution of this basis function increases considerably, coincides with the local maxima and onsets of plateaus in CASSCF(10,11)b.

For SA4-CASSCF(10,11)b, the ground state curve has a maximum at 3.0 \AA{} and strongly underestimated energies beyond that point. $S_1$ and $S_2$ state surfaces remain nearly unchanged, while $S_3$ now features a gentle maximum at the same coordinate as the ground state. In SA7-CASSCF(10,11)b, the ground state no longer has a local maximum, but all three excited states do. Simultaneously, they have significantly higher excitation energies throughout the entire PES, but more so near the FC region. The states $S_4$, $S_5$, $S_6$ (not shown) are all of 3$d\rightarrow$3d$_{x^2}$ character at extended bond lengths (they are strongly mixed at the FC region) and all feature a local minimum at the same geometry as the maxima of lower states. This method clearly captures the avoided crossing, but is, as a whole, of much worse quality. These characteristics are augmented further in SA12-CASSCF(10,11)b, where all excited states are dissociative and have higher excitation energies. Our attempts at explicitly incorporating the avoided crossing have brought behaviour that's more different from MRCI/CASSCF(6,17), and at a greater computational cost. The smaller active space, SA4-CASSCF(6,7), produces the most quantitatively agreeable behaviour, and so it was chosen to be used in further dynamic calculations.

To complete static studies, PES for triplet states were calculated at SA3-CASSCF(6,7) level of theory (see Figure S3d). Calculated triplets are the analogues to the excited states of SA4-CASSCF(6,7). $T_1$ and $T_2$ are almost coincident in energy outside of the FC region. The character of triplet states, assigned based on contributing orbital transitions, is discontinuous around 2.00 -- 2.75 \AA{}. This suggests the existence of an accessible conical intersection seam between 1$^3$A$_1$ and 1$^3$B$_1$, or $T_3$ and $T_2$ states. This would enable quick depopulation of the T$_3$ state, and support the observation of only two emissive triplet states~\cite{Farrell1999}. The energy barrier of both $T_1$ and $T_2$ is under 0.1 eV, which is lower than for singlet states.

PES singlet curves of Cr--CO$_{\text{eq}}$ were calculated using SA4-CASSCF(6,7) (see Figure S4). Ground state in this coordinate has a higher energy barrier of 1.2 eV and excited state barriers of 0.7 -- 1.0 eV. Furthermore, $\pi^*_{\text{bpy}}$ orbital gains no contribution from the 3$d_{x^2}$ orbital throughout the curve, due to its perpendicular position to the CO$_{\text{eq}}$ ligand. This directionality of the avoided crossing is consistent with ligand selectivity of photodissociation. Simultaneously, the excited energy barriers are low enough to allow for equatorial dissociation, if a complex is designed with an appropriate metal centre and ligands, which both stabilize axial bonds and destabilize equatorial bonds.

From a simple electrostatic picture, Cr$\rightarrow$bpy excitations lead to the depopulation of the metal centre. This leaves less electron density for bonding between Cr and CO. Consequently, \emph{any} Cr$\rightarrow$bpy excitation would have a photodissociative effect, regardless of spin, which is in line with CO being a spectator ligand of the excitation process. Indeed, the lowering of excited state energy barriers is observed in both CO$_{\text{ax}}$ and CO$_{\text{eq}}$, but the absence of 3$d_{x^2}$ avoided crossing results in bound shape of CO$_{\text{eq}}$ excited states.

\paragraph*{Electronic dynamics}
\begin{figure*}
    \centering
    \includegraphics[width=.8\linewidth]{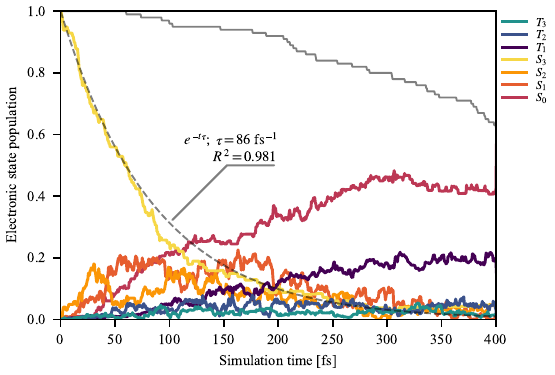}
    \caption{Time evolution of the electronic active state MCH populations. Trajectories which ended early stop contributing at the time of failure, and the ensemble is renormalized over the remaining trajectories. Gray solid line indicates the fraction of remaining trajectories. Gray dashed line is the monoexponential fit of $S_3$ decay over 300 fs.}
    \label{Fig:MCHActivePop}
\end{figure*}
    
The evolution of Molecular Coulomb Hamiltonian (MCH) active state electronic population is presented in Figure \ref{Fig:MCHActivePop}. The bright $S_3$ state starts decaying instantly and fits well with a monoexponential decay function with a lifetime of 86 fs (over 300 fs, R$^2$ = 0.981). This is in good agreement with the solvent-independent experimental value of 96 fs~\cite{Ma2016}. Electronic population of $T_3$ and $T_2$ never exceed 0.05 and 0.1, respectively. $T_1$ rises approximately linearly during 50 -- 300 fs, reaching a plateau of $\sim$0.19. Calculating triplet state lifetimes would require much longer simulation times, but the existence of a $T_1$ plateau does indicate a much slower decay to $S_0$, in accordance with their measured picosecond lifetimes~\cite{Farrell1999,Ma2016}. The ground state population increases in a semi-stepwise manner, with three flat regions at 0 fs, $\sim$120 fs and $\sim$300 fs. The first two regions suggest that transitions to the ground state are favoured at stretched bond geometries (discussed below), while the third is more likely due to the exhaustion of higher singlet state population and dephasing of coherent oscillations.

\begin{figure*}[t]
    \centering
    \includegraphics{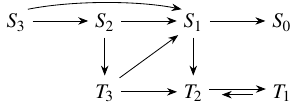}
    \caption{Proposed mechanism for the main deactivation pathway of \TheComplex.}
    \label{Fig:mechanism}
\end{figure*}

To complement electronic population data, the total count of hopping events, and the net difference of downwards and upwards hops are presented in Table \ref{Tab:hopcounts}. The resulting approximate mechanism is shown in Figure \ref{Fig:mechanism}. The main transfer of population goes through a cascade of $S_3\rightarrow S_2\rightarrow S_1\rightarrow S_0$, with branching points $S_2\rightarrow T_3$ and $S_1\rightarrow T_2$, and non-negligible direct transfer of $S_3\rightarrow S_1$. $S_3\rightarrow S_2$ transitions appear to occur mainly at the very beginning of the simulation, suggesting that this is the only allowed transition in the FC region. The 'middle' states $S_2$, $S_1$, $T_3$, and $T_2$ are in dynamic equilibrium, owing to low energy separation, but with a net flow towards $S_1$ and $T_2$. From there, trajectories decay either via $S_1\rightarrow S_0$ or $T_2\rightarrow T_1$. Transitions to $S_0$ are mostly irreversible, though some back hops are observed. The $S_0$ state energy is frequently close to the 'middle' manifold during a trajectory, but once a transition to $S_0$ occurs, ground and excited state energies quickly diverge, preventing back hops. This is true even if a dissociating trajectory drops to the ground state, suggesting that this is due to a nuclear rearrangement of the complex, fully coordinated or not. Finally, there is a dynamic equilibrium between $T_2$ and $T_1$, which is in agreement with the observed two trapping triplet states~\cite{Farrell1999,Vlcek2002}. The lack of direct transfer $S_{1-3}\rightarrow T_1$ is in accordance with previous observations~\cite{Ma2016}. The apparent equilibrium between $T_1$ and $S_0$, as judged by the number of hops in Table \ref{Tab:hopcounts}, is due to low energy gaps and state mixing on trajectories in the $T_1$ state. The low net transfer into $T_3$ and $T_2$ states does not allow for a confident confirmation of the existence of a $T_3/T_2$ conical intersection, which was suggested in context of calculated PES.

\begin{table*}[t]
    \begin{center}
        \input{Tab-hopcounts.tex}
    \end{center}
    \caption{Number of MCH hopping events within 300 fs. Transitions are from states in the first column to states in the first row. Top: total count; Bottom: net difference between downward and upward hops.}
    \label{Tab:hopcounts}
\end{table*}

\begin{figure}[ht]
    \centering
    \includegraphics[width=.8\linewidth]{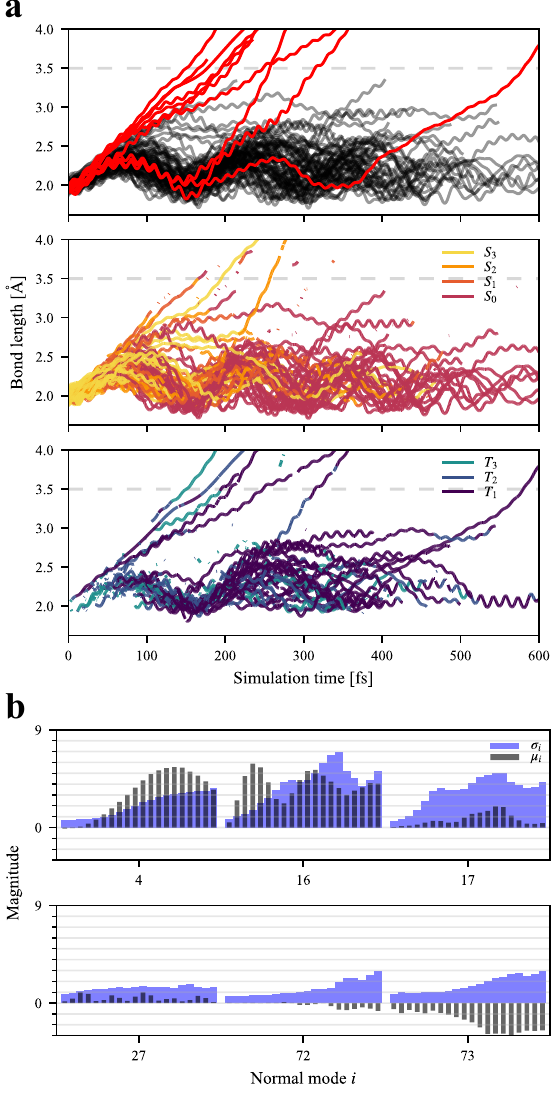}
    \caption{\textbf{a}: Evolution of Cr--CO$_{\text{ax}}$ bond length of all trajectories. Top: coloured red if a trajectory reached 3.5 \AA{} (threshold indicated by a dashed line). Middle and Bottom: coloured by currently occupied MCH state, split into singlet (middle) and triplet (bottom) states.
    \textbf{b}: Mean displacement $\mu_i$ and standard deviation $\sigma_i$ for selected mass-frequency scaled normal modes $i$ of the trajectory ensemble. Average values were taken over each 20 fs interval, up to the final value of 400 fs.}
    \label{Fig:BondLengths}
\end{figure}

\paragraph*{Nuclear dynamics}
For each trajectory, the evolution of the two equivalent Cr--CO$_{\text{ax}}$ bonds was tracked, and the one which has reached a greater value is shown in Figure \ref{Fig:BondLengths}a. Bond length of all Cr--CO$_{\text{eq}}$ is shown in Figure S5. No dissociation of an equatorial ligand was observed, in accordance with literature \cite{Farrell1999}. For all trajectories, if the Cr--CO$_{\text{ax}}$ bond length has passed 3.5 \AA{}, it was not observed to drop back below this value, so it was chosen as the threshold after which a trajectory is considered dissociated. Upon excitation to the $S_3$ state, all trajectories begin stretching their Cr--CO bonds. For CO$_{\text{ax}}$ (CO$_{\text{eq}}$), they begin at $\sim$2.02 (1.92) \AA{} and oscillate between 2.0 (1.9) and 2.3 (2.1) \AA{} with a period of $\sim$150 (80) fs. Upon reaching the first maximum of Cr--CO$_{\text{ax}}$, most trajectories proceed back down the oscillation pathway, while a small subset evolves towards longer bond lengths. Of this subset, some do not dissociate, instead taking a longer arc down, or hovering close to 3.0 \AA{} for over 100 fs. The remaining other trajectories proceed beyond this avoided crossing region and dissociate.

At 400 fs, 9 out of 70 trajectories were observed to dissociate (7 dissociated trajectories were stopped before 400 fs), and all did so as a continuation of the upward motion of the Cr--CO$_{\text{ax}}$ oscillation, giving a quantum yield of 13\%. It is in rough agreement with experimental work, where it ranges from 1\% to 10\%, depending on experimental conditions~\cite{Balk1980, Vichova1992}. Seven trajectories dissociated during the first period, and two more on the second period. One additional trajectory dissociated beyond 400 fs, on the third period.

To investigate the importance of other motions, Figure \ref{Fig:BondLengths}b presents the time evolution of selected frequency-mass scaled normal coordinates $\nu$. For each mode $i$, the mean displacement $\mu_i$ and standard deviation $\sigma_i$ (also referred to as \textit{activity}) are presented in grey and blue, respectively. These parameters were averaged in slices of 20 fs, up to 400 fs. Analysis of all modes is shown in Figure S6. Majority of normal modes show minimal temporal change. Those which vary most in time all involve carbonyl ligands: $\nu_{72}$ and $\nu_{73}$ are the antisymmetric and symmetric axial C-O stretching modes, $\nu_{16}$ and $\nu_{17}$ are symmetric and antisymmetric Cr-CO$_{\text{ax}}$ stretching modes, and $\nu_4$ involves $_{\text{ax}}$OC-Cr-CO$_{\text{ax}}$ bending in the $xz$-plane, with a lesser contribution of equatorial $_{\text{eq}}$OC-Cr-CO$_{\text{eq}}$ bending.
    
Several modes are set in motion in a coherent manner upon the initial electronic excitation: $\nu_{12}$ and $\nu_{14}$ which both have strong contribution from symmetric stretching of Cr-N bonds; $\nu_{22}$ the symmetric stretch of Cr-CO\textsubscript{eq}; $\nu_{19}$, $\nu_{35}$, $\nu_{38}$, $\nu_{45}$, $\nu_{60}$, $\nu_{64}$, $\nu_{66}$ which are all stretching modes of bipyridine bonds; $\nu_{27}$ which involves OC-Cr-CO and Cr-C-O bending of all carbonyl ligands. Of these, only modes $\nu_{22}$ and $\nu_{27}$ show a sizeable change in activity over time.

Several low frequency modes, in particular axial C-Cr-C bending $\nu_4$, increase in activity in a semi-stepwise manner and not immediately upon excitation. The onset of this sharp increase happens at around 100 fs, which is the same period of time at which the Cr-CO$_{\text{ax}}$ bond length was noted to diverge between dissociation, contraction, or hovering for an extended period. When considering dissociating trajectories alone (Insets of Figure S6), the activity of $\nu_{16}$ and $\nu_{17}$ is much greater, while the activity of $\nu_{4}$ is considerably reduced, though the mean displacement remains high in both. This behaviour suggests a coupling between Cr-CO$_{\text{ax}}$ stretching and $_{\text{ax}}$OC-Cr-CO$_{\text{ax}}$ bending modes, which would act to inhibit dissociation.

These observations suggest a ballistic model of the photodissociation process. Looking again at PES of Figure \ref{Fig:PES}, the $S_3$ state has a downward gradient away from the FC region at 2.02 \AA{}, which is the center of the initial Wigner distribution. This slope gives the system kinetic energy towards dissociation. For most trajectories, this is not enough to overcome the energy barrier: the bond contracts back. A smaller subset will have kinetic energy which is approximately equal to the barrier. These trajectories will hover in the avoided crossing region plateau, on top of the energy barrier, before turning back after an extended period. Finally, those with even more energy will overcome the barrier with sufficient leftover kinetic energy to continue the motion towards dissociation on a flat surface. This is supported by the observation, that the initial bond lengths of all dissociating trajectories are in the shorter half of the initial Wigner distribution. As time proceeds, for non-dissociating trajectories, the excess energy within this oscillating normal mode is expected to equipartition into other modes of motion, making the dissociation less likely after subsequent periods. In other words, if a trajectory did not dissociate during the first period of bond oscillation (around 150 fs), it is less likely to dissociate later on.
It is still possible as a probabilistic process, and three trajectories in total were observed to dissociate on second and third periods, but in significantly lesser numbers. It is also noted that solvent may prohibit dissociation after the first period of oscillation. A ballistic model is consistent with the $\sim$100 fs timescale of photodissociation~\cite{Farrell1999, Ma2016} and wavelength-dependence of the quantum yield, as occupying a higher vibrational state will directly favour dissociation.
The evolution of Cr-CO$_{\text{eq}}$ (Figure S5) shows no dissociation. When compared with Cr-CO$_{\text{ax}}$, shorter period of oscillation (80 vs. 150 fs) and shorter bond length at minimum geometry (1.92 vs 2.02 \AA{}) indicate a stronger bond, consistent with the bound shape of PES and higher energy barries (Figure S4).
Slight overestimation of the photodissociation quantum yield is a possible consequence of neglecting solvent effects. An explicit solvent could act as a physical barrier for the released molecule, shifting the energy barrier upwards, as well as damping vibrational motion, prohibiting dissociation beyond the first period of Cr--CO$_{\text{ax}}$ oscillation.

\paragraph*{Combined Nuclear and Electronic Dynamics}
In this section, we combine the electronic and nuclear dynamic information, seen on middle and bottom panels of Figure \ref{Fig:BondLengths}a. Seven out of ten dissociating trajectories are of triplet character at 3.5 \AA{}. Five of these trajectories undergo intersystem crossing long before reaching the energy barrier at $\sim$3.0 \AA{}; the remaining two do so in its vicinity. This is in contrast to the established understanding that triplet states are non-dissociative. It is however consistent with the calculated PES, wherein $T_1$ and $T_2$ have lower energy barriers of dissociation than all singlet states. Therefore, if a trajectory undergoes intersystem crossing early on, it will approach a lower energy barrier and be more likely to cross it.
This supports the above explained mechanism, where the main determining factor is the energy barrier, and not the occupied electronic and spin state.

\begin{figure}
    \centering
    \includegraphics[width=.9\linewidth]{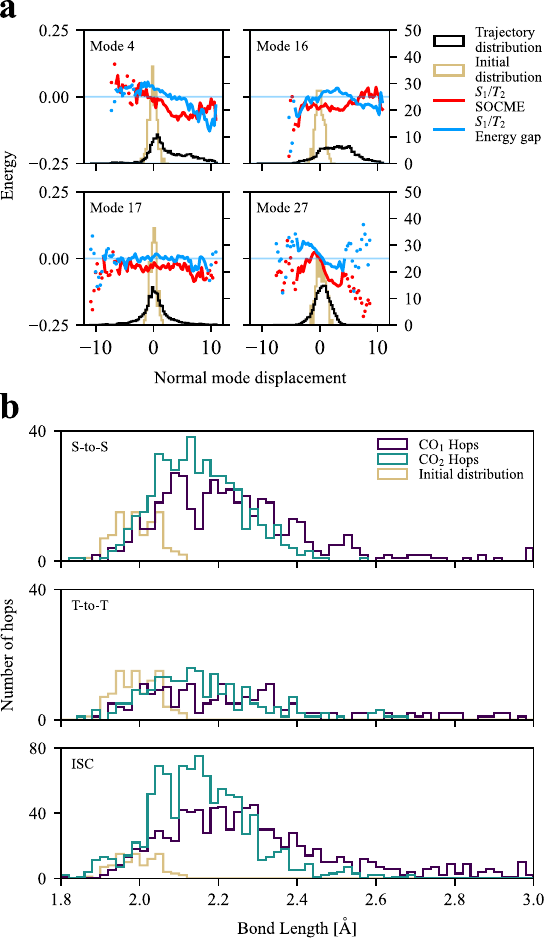}
    \caption{\textbf{a}: Energy gap and spin-orbit coupling matrix elements of $S_1$/$T_2$ crossing, averaged over the nuclear displacement of selected normal modes. Distribution of all geometries visited during dynamics, and the initial Wigner distribution are given for reference. Energy gap values are given in eV (left axis), SOCME are given in cm$^{-1}$ (right axis). Areas with under 100 samples are indicated with dots; \textbf{b}: Histograms of all observed hops as a function of Cr--CO$_{\text{ax}}$ bond length. For each trajectory, the farther reaching CO$_{\text{ax}}$ is denoted CO$_1$, and the other CO$_2$. Initial distribution is shown in red.}
    \label{Fig:HopHistogramsSOCdist}
\end{figure}

A remaining question is how the nuclear conformation affects the probability of transitions. Cannizzo \emph{et al.}~\cite{Cannizzo2008} have noted a correlation between metal-ligand vibrational frequencies and inter-system crossing time scales for analogous systems Re(L)(CO)$_3$(bpy) [L = Cl, Br, I]. This suggests that a distortion along the ligand-metal stretching mode leads to an area with greater spin-orbit coupling and more allowed intersystem crossings. Similar observations were noted for the lifetimes of singlet MLCT states in Fe and Ru tris-bpy complexes~\cite{Cannizzo2006,Gawelda2007}. In present simulations, the semi-stepwise increase of $S_0$ population points towards a similar behaviour.

To investigate the coupling between nuclear motions and electronic spin transitions, Figure \ref{Fig:HopHistogramsSOCdist}a presents: (1) energy gap between states $S_1$ and $T_2$; (2) norm of the spin-orbit coupling matrix element (SOCME), averaged over all trajectories as a function of normal mode displacement, for selected modes. A full plot over all modes is presented in Figure S7. We focus on the $S_1$/$T_2$ crossing, as it accounts for over half of all ISC hops (Table \ref{Tab:hopcounts}). Variation of SOCME was not observed to be large enough to indicate a favourable ISC at particular geometries. All states in the middle manifold ($S_{1-2}$ and $T_{1-3}$) were observed to evolve in parallel, that is the pairwise energy gap remained near constant. For the $S_1$/$T_2$ pair, the energy gap was very close to degeneracy, as can seen most clearly for mode $\nu_{17}$. A weak dependence of energy gap on nuclear displacement was noted for various bending modes involving Cr-L bonds, with example modes $\nu_{4}$ and $\nu_{27}$ being shown. The symmetric Cr-CO$_{\text{ax}}$ stretching mode does also show a weak dependence, and with the previously shown coupling with $\nu_{4}$, they would account for a weakly favoured ISC at extended Cr-CO$_{\text{ax}}$ geometries. However, the near-degeneracy of these two states remains as the dominant driving force for ISC.
Figure \ref{Fig:HopHistogramsSOCdist}b shows distributions of all hops and the initial Wigner distribution as a function of Cr--CO$_{\text{ax}}$ bond length.
For each trajectory, hops are shown against bond lengths of both CO$_{\text{ax}}$, with the one which has reached the farthest denoted CO$_1$; the other CO$_2$. CO$_2$ acts as a control distribution, approximating one that would statistically arise from vibrational motion alone.
For CO$_1$ (CO$_2$), median of initial distribution is 1.99 \AA{} while the distributions of hops have medians of 2.21, 2.24 and 2.27 \AA{} (2.14, 2.12 and 2.14 \AA{}) for $S\rightarrow S$, $T\rightarrow T$ and intersystem crossing hops, respectively. Comparison of CO$_1$ and CO$_2$ distributions shows that ISC hops are weakly favoured at stretched bond geometries, in accordance with the above analysis of energy gap dependence.

%% file: Tab-hopcounts.tex
    \begin{tabular}{r|*{4}{w{c}{.5cm}}|*{3}{w{c}{.5cm}}}
     $\nearrow$ & $S_0$ & $S_1$ & $S_2$ & $S_3$ & $T_1$ & $T_2$ & $T_3$ \\ \toprule
          $S_0$ &   -   &   16  &   1   &   4   &   64  &   6   &   0   \\ 
          $S_1$ &   57  &   -   &   85  &   3   &   8   &  235  &   50  \\
          $S_2$ &   8   &  121  &   -   &   11  &   0   &   2   &   62  \\
          $S_3$ &   12  &   34  &   62  &   -   &   0   &   1   &   5   \\ \midrule
          $T_1$ &   64  &   3   &   0   &   0   &   -   &   42  &   5   \\
          $T_2$ &   6   &  215  &   3   &   0   &   61  &   -   &   17  \\
          $T_3$ &   0   &   56  &   54  &   0   &   4   &   22  &   -   \\ \bottomrule
    \end{tabular}
\\ \medskip
    \begin{tabular}{r|*{4}{w{c}{.5cm}}|*{3}{w{c}{.5cm}}}
        $\nearrow$ & $S_0$ & $S_1$ & $S_2$ & $S_3$ & $T_1$ & $T_2$ & $T_3$ \\ \toprule
        $S_0$ &   -   &       &       &       &       &       &       \\ 
        $S_1$ &   41  &   -   &       &       &   5   &   20  &       \\
        $S_2$ &   7   &   36  &   -   &       &       &       &   8   \\
        $S_3$ &   8   &   31  &   51  &   -   &       &   1   &   5   \\ \midrule
        $T_1$ &       &       &       &       &   -   &       &   1   \\
        $T_2$ &       &       &   1   &       &   19  &   -   &       \\
        $T_3$ &       &   6   &       &       &       &   5   &   -   \\ \bottomrule
    \end{tabular}

%% file: Conclusions.tex
We have theoretically investigated the photochemistry of \TheComplex{} with surface hopping non-adiabatic dynamics in combination with CASSCF for the electronic structure. The calculated photodissociation quantum yield of an axial CO ligand is 13\% and the predicted lifetime of $S_3$ bright state is 86 fs, both in good agreement with experimental data~\cite{Balk1980,Vichova1992,Ma2016}. The initial excited state decays to a middle manifold of $S_2$, $S_1$, $T_3$ and $T_2$ electronic states, which are in fast dynamic equilibrium, owing to small state energy differences, with a net flow towards $S_1$ and $T_2$. From there, the system decays to the ground state via $S_1\rightarrow S_0$ transition, or is trapped in a dynamic equilibrium $T_2\leftrightarrow T_1$. Flow of electronic population into the triplet equilibrium, weakly influenced by bending modes around the central chromium atom, is in direct accordance with experimental presence of triplet trapping states. The present simulations show involvement of $S_1$ and $S_2$ states, which were not observed previously. The simulations show that the initial relaxation of the $S_3$ bright state is not a competitive process between photodissociation and intersystem crossing. Photodissociation is instead a ballistic process which can occur from both singlet and triplet states. It happens within 150 fs as the kinetic energy along the reactive mode of motion quickly dissipates into other modes. Equatorial carbonyls are not observed to dissociate, having higher energy dissociation barriers due to a lack of an avoided crossing with higher lying 3$d_{x^2}$ states.

%% file: Acknowledgements.tex
This work is funded by European Union (ERC 101040356 - ATTOP) and \emph{R\'{e}gion des Pays de la Loire}. Views and opinions expressed are however those of the authors only and do not necessarily reflect those of the European Union or the European Research Council Executive Agency. Neither the European Union nor the granting authority can be held responsible for them.
This work received financial support from the State through the EUR LUMOMAT project and the Investissements d’Avenir program ANR-18-EURE-0012.
Calculations were performed at Centre de Calcul Intensif des Pays de la Loire, GLiCID, located in Nantes. The authors are thankful to Denis Jacquemin for fruitful discussions, and to the Institute for Development and Resources in Intensive Scientific Computing (IDRIS) for allocation of computational time.

%% file: Tab-FC-benchmark.tex
\begin{sidewaystable*}
    \centering
	\begin{threeparttable}[b]
		\caption[FC Benchmark of electronic structure methods]{Transition energies of \TheComplex{} at the Frank-Condon geometry. Left side contains literature data, right side has CASSCF and TD-DFT data from this work.}
		\label{Tab:FC-Benchmark-SI}
		\begin{filecontents*}{anc/data.csv}
symmetry,tran1,tran2,empty column,cas1011,caspt1011,cas1012,caspt1012,cas617,mrci617,cas67,cas1011bpy,sa41011,sa71011,sa121011,pbe1,pbe2,pbe3,pbe4,b3lyp1,b3lyp2,b3lyp3,b3lyp4,bhhlyp1,bhhlyp2,bhhlyp3,bhhlyp4,cam1,cam2,cam3,cam4
A$_{1}$,$d_{xz}$,$\pi^*_{\text{bpy}}$,,2.799,2.18,,,3.014,2.692,2.33,2.432,2.278,2.855,3.23,2.331,2.303,2.37,2.355,2.24,2.202,2.269,2.249,2.818,2.784,2.883,2.86,2.637,2.604,2.709,2.685
,$d_{z^2-y^2}$,$\pi^*_{\text{CO}}$,,,,,,3.843,,,,,,,3.6,,,,3.467,,,,3.747,,,3.927,3.711,,,3.735
,$d_{xy}$,$\pi^*_{\text{bpy}}$,,,,3.561,3.145,4.105,3.54,,3.999,,,,3.358,,,,3.215,3.169,,,4.105,,,4.115,3.858,,,3.893
,$d_{xz}$,$\pi^*_{\text{bpy}}$,,,,,,4.166,3.809,,3.751,,,,3.077,3.037,3.103,3.078,2.919,2.87,2.939,2.908,3.936,3.726,3.769,3.746,3.623,3.588,3.66,3.631
,$d_{z^2-y^2}$,$\pi^*_{\text{CO}}$,,,,,,4.711,,,,,,,,,,,,,,,4.705,,,,4.708,,,
,$d_{xy}$,$\pi^*_{\text{CO}}$,,,,,,4.917,,,,,,,,,,,,,,,4.351,,,,,,,4.431
,$d_{z^2-y^2}$,$d_{x^2}$,,5.31,4.632,,,,,,,,,5.534,,,,,,,,,5.391,,,,4.526,,,
,$d_{xz}$,$\pi^*_{\text{CO}}$,,,,,,,,,,,,,,,,,,,,,,,,,,,,4.539
A$_{2}$,$d_{xy}$,$\pi^*_{\text{CO}}$,,,,,,3.211,,,,,,,3.53,,,,3.441,,,,3.587,3.652,3.649,3.647,3.504,3.585,3.566,3.572
,$d_{xz}$,$\pi^*_{\text{CO}}$,,,,,,3.313,,,,,,,3.201,3.244,3.228,3.229,3.151,,3.178,3.18,3.264,3.314,3.303,3.303,3.158,3.215,3.198,3.203
,$d_{z^2-y^2}$,$\pi^*_{\text{bpy}}$,,,,,,3.943,3.339,,3.801,,,,3.043,3.027,3.111,3.088,2.85,2.824,2.911,2.882,4.082,,,4.09,3.703,,,3.757
B$_{1}$,$d_{z^2-y^2}$,$\pi^*_{\text{bpy}}$,,2.356,1.566,,,2.451,2.145,1.984,2.029,1.901,2.607,2.951,1.855,1.844,1.944,1.919,1.729,1.708,1.808,1.777,2.616,2.589,2.714,2.685,2.327,2.311,2.45,2.415
,$d_{xy}$,$\pi^*_{\text{CO}}$,,,,,,3.464,,,,,,,3.316,,,,3.254,,,,3.357,3.41,3.41,3.411,3.267,3.331,3.323,3.33
,$d_{z^2-y^2}$,$\pi^*_{\text{bpy}}$,,,,,,3.73,3.682,,3.591,,,,2.89,2.841,2.902,2.873,2.698,2.639,2.702,2.665,3.998,3.73,3.769,3.968,3.629,3.562,3.644,3.604
,$d_{z^2-y^2}$,$\pi^*_{\text{CO}}$,,,,,,4.209,,,,,,,,,,,,,,,3.746,,,3.753,3.998,,,4.034
,$d_{xy}$,$d_{yz}$,,4.778,3.996,,,,,,,,,4.92,,,,,,,,,5.605,,,,5.384,,,
,$d_{xz}$,$d_{x^2}$,,5.323,4.503,,,,,,,,,5.37,,,,,,,,,5.25,,,,,,,
,$d_{z^2-y^2}$,$\pi^*_{\text{CO}}$,,,,,,,,,,,,,,,,,,,,,4.192,,,,4.237,,,4.255
B$_{2}$,$d_{xy}$,$\pi^*_{\text{bpy}}$,,2.255,1.792,,,2.746,2.266,2.101,2.224,2.035,2.699,3.125,1.984,1.975,2.076,2.06,1.878,1.857,1.961,1.939,2.563,2.544,2.663,2.641,2.379,2.363,2.5,2.476
,$d_{z^2-y^2}$,$\pi^*_{\text{CO}}$,,,,,,3.422,,,,,,,3.033,3.071,3.071,3.064,2.966,3.002,3.007,3.001,3.086,3.132,3.13,3.125,3.044,3.105,3.096,3.095
,$d_{xz}$,$\pi^*_{\text{bpy}}$,,,,3.551,3.193,4.334,3.666,,3.936,,,,3.279,3.255,3.324,3.306,3.138,3.117,3.172,3.151,4.012,,,4.039,3.778,,,3.823
,$d_{xy}$,$\pi^*_{\text{bpy}}$,,,,,,3.951,3.843,,3.74,,,,2.996,2.957,3.018,2.999,2.833,2.776,2.845,2.815,3.801,3.737,3.791,3.767,3.651,3.587,3.669,3.638
,$d_{z^2-y^2}$,$d_{yz}$ ,,4.239,3.676,,,,,,,,4.419,4.395,,,,,4.461,,,,4.902,,,,4.62,,,4.612
,$\pi_{\text{bpy}}$,$\pi^*_{\text{bpy}}$,,,,,,,,,,,,,,,,,4.724,,,,5.09,,,,,,,4.875
,$d_{xy}$,$d_{xz}$,,,,,,,,,6.006,,5.018,6.046,,,,,,,,,4.208,,,,4.396,,,
\end{filecontents*}
\pgfplotstabletypeset[fixed,precision=3,fixed zerofill,
        x = coord,
        col sep=comma,
        column type={c},
        every odd row/.append style={before row={\rowcolor[gray]{0.95}}},
        every col no 9/.style={column type={p{0mm}}},
        every last row/.style={after row=\cmidrule{1-9}\cmidrule{11-19}},
        every row no 7/.append style={after row=\cmidrule{1-9}\cmidrule{11-19}},
        every row no 10/.append style={after row=\cmidrule{1-9}\cmidrule{11-19}},
        every row no 17/.append style={after row=\cmidrule{1-9}\cmidrule{11-19}},
		columns={symmetry,tran1,tran2,cas1011,caspt1011,cas1012,caspt1012,cas617,mrci617,empty            column,cas67,cas1011bpy,sa41011,sa71011,sa121011,b3lyp1,pbe1,bhhlyp1,cam1},
		every head row/.style={output empty row,after row={
				\cmidrule{1-9}\cmidrule{11-19}
                &\multicolumn{2}{r}{Active Space:}&\multicolumn{2}{c}{(10,11)}&
                \multicolumn{2}{c}{(10,12)}&\multicolumn{2}{c}{(6,17)}
                &&
                (6,7)&(10,11)a&\multicolumn{3}{c}{(10,11)b}&&&&
				\\
				\cmidrule(lr){4-5}\cmidrule(lr){6-7}
                \cmidrule(lr){8-9}
                \cmidrule(lr){11-11}
                \cmidrule(lr){12-12}\cmidrule(lr){13-15}
				&\multicolumn{2}{r}{Method:}&CASSCF\tnote{a}&
                CASPT2\tnote{a}&CASSCF\tnote{a}&
                CASPT2\tnote{a}&CASSCF\tnote{b}&
                MRCI\tnote{b}&&SA4&SA12&SA4&SA7&SA12&
				\multicolumn{1}{c}{B3LYP}&\multicolumn{1}{c}{PBE0}&
                \multicolumn{1}{c}{BHHLYP}&\multicolumn{1}{c}{CAM-B3LYP}\\
                \cmidrule{1-9}\cmidrule{11-19}
            }}
		]{anc/data.csv}
		\begin{tablenotes}
			\item Basis set used for DFT calculations is 6-31G*, ANO-RCC-VDZ for CASSCF, all values in eV, DFT calculations were performed on B3LYP/6-31G$^*$ geometries;
			\item[a] From Ref. \cite{Guillaumont1997};
			\item[b] From Ref. \cite{Guillaumont2001};
		\end{tablenotes}
	\end{threeparttable}
\end{sidewaystable*}

\begin{sidewaystable*}
    \centering
	\begin{threeparttable}
		\caption[FC Benchmark of TD-DFT basis set dependence]{Transition energies of \TheComplex{} calculated using four DFT functionals and four basis sets on B3LYP/6-31G$^*$ geometry.}
		\label{Tab:FC-Benchmark-2-SI}
		\rowcolors{3}{gray!10}{white}
		\pgfplotstabletypeset[fixed,precision=3,fixed zerofill,
        x = coord,
        col sep=comma,
        column type={c},
        every last row/.style={after row=\bottomrule},
        every row no 7/.append style={after row=\midrule},
        every row no 10/.append style={after row=\midrule},
        every row no 17/.append style={after row=\midrule},
		columns={symmetry,tran1,tran2,b3lyp1,b3lyp2,b3lyp3,b3lyp4,
			     pbe1,pbe2,pbe3,pbe4,bhhlyp1,bhhlyp2,bhhlyp3,bhhlyp4,cam1,cam2,cam3,cam4
		},
		every head row/.style={output empty row,after row={
				\toprule&\multicolumn{2}{c}{}&\multicolumn{4}{c}{B3LYP}&
                \multicolumn{4}{c}{PBE0}&\multicolumn{4}{c}{BHHLYP}&
                \multicolumn{4}{c}{CAM-B3LYP}\\
				\cmidrule(lr){4-7}\cmidrule(lr){8-11}\cmidrule(lr){12-15}\cmidrule(lr){16-19}
				&\multicolumn{2}{c}{Transition}&
				\Romannum{1}&\Romannum{2}&\Romannum{3}&\Romannum{4}&
				\Romannum{1}&\Romannum{2}&\Romannum{3}&\Romannum{4}&
				\Romannum{1}&\Romannum{2}&\Romannum{3}&\Romannum{4}&
				\Romannum{1}&\Romannum{2}&\Romannum{3}&\Romannum{4}\\\midrule}}
		]{anc/data.csv}
		\begin{tablenotes}
			\item Basis sets used are: (\Romannum{1}) 6-31G$^*$; (\Romannum{2}) 6-311G$^*$; (\Romannum{3}) 6-31+G$^*$; (\Romannum{4}) 6-311+G$^*$; all values in eV;
		\end{tablenotes}
	\end{threeparttable}
\end{sidewaystable*}

%% file: Tab-FC-B3LYP-xyz.tex
\begin{table}
    \centering
    \caption[B3LYP geometry of \TheComplex{}]{Geometry of \TheComplex{} structure optimized at the B3LYP/6-31G* level of theory, under $C_{2v}$ symmetry constraint. All units in \AA{}.}
    \label{Tab:B3LYP-geom}
    \begin{tabular}{c*{3}{D{.}{.}{8}}}
        Atom&\multicolumn{1}{c}{$x$} & \multicolumn{1}{c}{$y$} & \multicolumn{1}{c}{$z$} \\\toprule
        Cr &  0.00000000  &  0.00000000  & 1.09152833 \\
        C  & -1.90097191  &  0.00000000  & 1.17491192 \\
        C  &  1.90097191  &  0.00000000  & 1.17491192 \\
        C  &  0.00000000  &  1.32631596  & 2.36595293 \\
        C  &  0.00000000  & -1.32631596  & 2.36595293 \\
        O  &  0.00000000  &  2.18497353  & 3.15290140 \\
        O  &  0.00000000  & -2.18497353  & 3.15290140 \\
        O  & -3.05294832  &  0.00000000  & 1.28580416 \\
        O  &  3.05294832  &  0.00000000  & 1.28580416 \\
        C  &  0.00000000  &  2.65000899  &-0.49438931 \\
        C  &  0.00000000  & -2.65000899  &-0.49438931 \\
        C  &  0.00000000  &  1.51317415  &-2.97647707 \\
        C  &  0.00000000  & -1.51317415  &-2.97647707 \\
        C  &  0.00000000  &  3.47898973  &-1.60882677 \\
        C  &  0.00000000  & -3.47898973  &-1.60882677 \\
        H  &  0.00000000  &  3.06055570  & 0.50866552 \\
        H  &  0.00000000  & -3.06055570  & 0.50866552 \\
        H  &  0.00000000  &  1.03902960  &-3.95082853 \\
        H  &  0.00000000  & -1.03902960  &-3.95082853 \\
        H  &  0.00000000  &  4.55664385  &-1.48225739 \\
        H  &  0.00000000  & -4.55664385  &-1.48225739 \\
        C  &  0.00000000  &  2.89865636  &-2.87780513 \\
        C  &  0.00000000  & -2.89865636  &-2.87780513 \\
        H  &  0.00000000  &  3.51346060  &-3.77216446 \\
        H  &  0.00000000  & -3.51346060  &-3.77216446 \\
        C  &  0.00000000  & -0.73570148  &-1.81258473 \\
        C  &  0.00000000  &  0.73570148  &-1.81258473 \\
        N  &  0.00000000  &  1.30427946  &-0.57793451 \\
        N  &  0.00000000  & -1.30427946  &-0.57793451
    \end{tabular}
\end{table}

%% file: Tab-dx2-contribution-SI.tex
\begin{tabular}{ll*{6}{D{.}{.}{4}}}
    Distance (\AA) &      &  \multicolumn{1}{c}{1.903}   &  \multicolumn{1}{c}{2.1}   &  \multicolumn{1}{c}{2.4}   &  \multicolumn{1}{c}{3.0}   &  \multicolumn{1}{c}{4.0}   &  \multicolumn{1}{c}{5.0} \\\toprule
    SA4-CAS(6,7)     & CO$_{\text{ax}}$ & 0.0000 & 0.0174 & 0.0409 & 0.0748 & 0.0995 & 0.1050 \\
                     & CO$_{\text{eq}}$ & 0.0000 & 0.0000 & 0.0000 & 0.0000 & 0.0000 & 0.0001 \\
    SA4-CAS(10,11)b  & CO$_{\text{ax}}$ & 0.0000 & 0.0312 & 0.0728 & 0.1548 & 0.3741 & 0.4374 \\
    SA7-CAS(10,11)b  & CO$_{\text{ax}}$ & 0.0000 & 0.0551 & 0.3589 & 0.5417 & 0.5348 & 0.5375 \\
    SA12-CAS(10,11)b & CO$_{\text{ax}}$ & 0.0000 & 0.0188 & 0.5682 & 0.5421 & 0.5304 & 0.5269 \\\bottomrule
\end{tabular}

%% file: Tab-FC-CASSCF67-xyz.tex
\begin{table}
    \centering
    \caption[CASSCF(6,7) geometry of \TheComplex]{Geometry of \TheComplex{} structure optimized at the CASSCF(6,7)/ANO-RCC-VDZ level of theory. All units in \AA{}.}
    \begin{tabular}{c*{3}{D{.}{.}{8}}}
        Atom&\multicolumn{1}{c}{$x$} & \multicolumn{1}{c}{$y$} & \multicolumn{1}{c}{$z$} \\\toprule
        Cr &  0.00000325 & -0.00003401 & 1.12178328 \\
        C  & -2.00433708 &  0.00001346 & 1.18525461 \\
        C  &  2.00434504 & -0.00004453 & 1.18523705 \\
        C  &  0.00003315 &  1.37877261 & 2.46070836 \\
        C  & -0.00001155 & -1.37887055 & 2.46066272 \\
        O  &  0.00003655 &  2.19545629 & 3.28051239 \\
        O  & -0.00001705 & -2.19558290 & 3.28044065 \\
        O  & -3.14271240 &  0.00003567 & 1.28000614 \\
        O  &  3.14271990 & -0.00003131 & 1.27999552 \\
        C  &  0.00000273 &  2.65195780 & 0.52822297 \\
        C  & -0.00003633 & -2.65190431 & 0.52819697 \\
        C  & -0.00005912 &  1.50689808 & 2.98937965 \\
        C  &  0.00001557 & -1.50688914 & 2.98936366 \\
        C  & -0.00001690 &  3.46616136 & 1.63312433 \\
        C  &  0.00000142 & -3.46613077 & 1.63307239 \\
        H  &  0.00007339 &  3.06615916 & 0.45972775 \\
        H  & -0.00004311 & -3.06608377 & 0.45976141 \\
        H  & -0.00016173 &  1.04156094 & 3.95494522 \\
        H  & -0.00006829 & -1.04157948 & 3.95494203 \\
        H  & -0.00010635 &  4.53414507 & 1.51215097 \\
        H  &  0.00000858 & -4.53411163 & 1.51207070 \\
        C  & -0.00017400 &  2.88189321 & 2.89543659 \\
        C  &  0.00002724 & -2.88188254 & 2.89539706 \\
        H  &  0.00031138 &  3.48967310 & 3.78321785 \\
        H  &  0.00020371 & -3.48967770 & 3.78316758 \\
        C  & -0.00001840 & -0.73785636 & 1.83264273 \\
        C  & -0.00002009 &  0.73789712 & 1.83264291 \\
        N  &  0.00002062 &  1.30794824 & 0.61732767 \\
        N  & -0.00002014 & -1.30789313 & 0.61732421
    \end{tabular}
\end{table}